\documentclass[twocolumn,twoside]{IEEEtran}
\usepackage{graphicx}

\begin{document}

\title{IPTV Over ICN}

\author{
\IEEEauthorblockN{George Xylomenos\IEEEauthorrefmark{1}, 
Alexander Phinikarides\IEEEauthorrefmark{2},\\
Ioannis Doumanis\IEEEauthorrefmark{3},
Xenofon Vasilakos\IEEEauthorrefmark{1},
Yannis Thomas\IEEEauthorrefmark{1}
Dirk Trossen\IEEEauthorrefmark{4},
Michael Georgiades\IEEEauthorrefmark{2},
Stuart Porter\IEEEauthorrefmark{3}
}\\
\IEEEauthorblockA{\IEEEauthorrefmark{1}Athens University of Economics and Business, Greece}
\IEEEauthorblockA{\IEEEauthorrefmark{2}PrimeTel PLC, Cyprus}
\IEEEauthorblockA{\IEEEauthorrefmark{3}CTVC Ltd, UK}\\
\IEEEauthorblockA{\IEEEauthorrefmark{4}InterDigital Europe, UK}
}
 \maketitle

\begin{abstract}
The efficient provision of IPTV services requires support for IP multicasting and IGMP snooping, limiting such services to single operator networks. Information-Centric Networking (ICN), with its native support for multicast seems ideal for such services, but it requires operators and users to overhaul their networks and applications. The POINT project has proposed a hybrid, IP-over-ICN, architecture, preserving IP devices and applications at the edge, but interconnecting them via an SDN-based ICN core. This allows individual operators to exploit the benefits of ICN, without expecting the rest of the Internet to change. In this paper, we first outline the POINT approach and show how it can handle multicast-based IPTV services in a more efficient and resilient manner than IP. We then describe a successful trial of the POINT prototype in a production network, where real users tested actual IPTV services over both IP and POINT under regular and exceptional conditions. Results from the trial show that the POINT prototype matched or improved upon the services offered via plain IP.
\end{abstract}


\section{Introduction}

Information-Centric Networking~(ICN)~\cite{icn_survey} proposes replacing the endpoint-based communication of the current Internet with an architecture focusing on the exchange of named data. An active research community has grown around ICN, following different design approaches, but as ICN requires overhauling the entire Internet infrastructure \emph{and} rewriting all applications, it is hard to make the transition from research testbeds to operational networks. 

Based on their experience in pioneering ICN projects, including PSIRP, PURSUIT and COMET, the partners of the POINT project have long realized the difficulty of replacing a hugely successful incumbent architecture with a promising, but unproven, clean-slate one. Therefore, POINT has taken a more pragmatic approach to the introduction of ICN: instead or replacing the Internet wholesale, POINT proposes supporting the existing IP-based applications and services over an ICN core network~\cite{eucnc}. POINT thus addresses the needs of individual operators who want to retain compatibility with the existing Internet, while taking advantage of specific ICN solutions to improve the performance of their network. 

One specific case study of such potential improvements can be found in IPTV services. To support such services efficiently, operators must resort to non-standard extensions to support IP multicast, such as IGMP snooping (explained in Section~\ref{snoop}), effectively limiting such services to single operator networks. With POINT and its native multicast capability, IPTV applications can be supported with plain SDN switches, while being more flexible in terms of traffic handling under regular and exceptional conditions.

This paper first provides an overview of the POINT approach. Then, it describes how IPTV is implemented in the current Internet, and how it is supported with POINT. We then describe a successful trial of the POINT technology over the network of project partner PrimeTel, with real users accessing an IPTV service over either IP or POINT. We show how POINT was able to improve upon IP under exceptional conditions, based on data gathered during the trial. 

\section{The POINT architecture}

The POINT architecture aims to replace the network of an individual network operator, so as to improve its IP-based services. POINT is a drop-in replacement for the existing network: it does not require changes to the existing IP User Equipment~(UE), or to the IP routers/gateways of interconnected operators. This is achieved by combining an ICN core network with a set of Network Attachment Points~(NAPs) which serve as gateways between IP and ICN.

The baseline POINT architecture was derived from the PURSUIT ICN architecture~\cite{PURSUIT}, in which the UEs can publish and subscribe to named information items. This publish/subscribe architecture is facilitated by three core functions: a Rendezvous~(RV) function that matches publisher and subscriber nodes; a Topology Management~(TM) function that calculates paths between the various nodes and encodes them into Forwarding Identifiers~(FIDs); and, a Forwarding Node~(FN) function that allows data items to be forwarded in the network based on the FIDs. The FIDs represent the set of links that a packet must traverse, whether these form a unicast path or a multicast tree. These FIDs are included in packet headers, allowing FNs to forward packets with a few bitwise operations, without requiring routing tables or other state. Consequently, POINT enables stateless multicast switching and native anycast.
 
\begin{figure}
\centering
\includegraphics[width=3.25in]{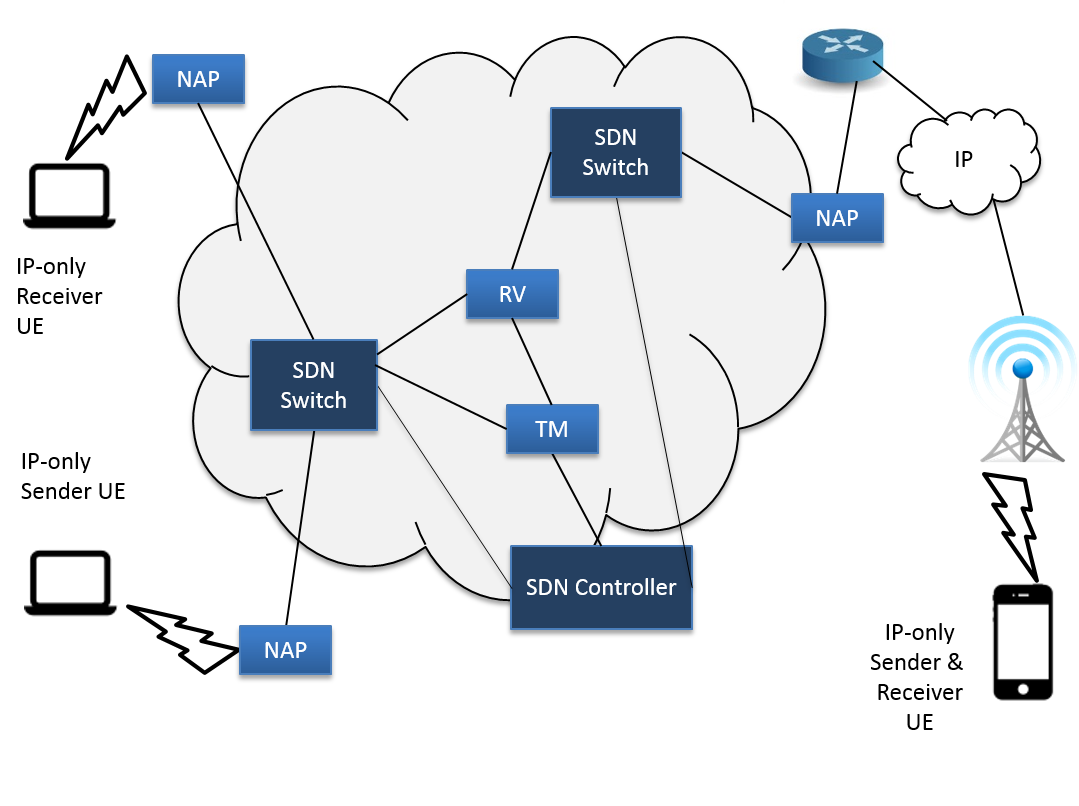}
\caption{The POINT architecture.}\label{architecture}
\end{figure}

Figure~\ref{architecture} outlines the main components of the POINT architecture, showing the physical connections between the various entities. The RV and TM functionalities, often combined in a single Path Computation Entity~(PCE) function, are the main control functions of the ICN cloud. Standard Software Defined Networking~(SDN) switches are used for the FN functionality in POINT, replacing the dedicated forwarding nodes of PURSUIT~\cite{SDN}. The SDN switches are unaware of POINT, therefore they are controlled by an SDN controller~\cite{odl}, which communicates with the TM function: the TM instructs the SDN controller how to configure the SDN switches so as to translate the FIDs included in packets to forwarding actions on their attached links, while the SDN controller informs the TM of any changes in the topology and operation of the network.

Another new feature introduced by POINT is the fast formation of multicast trees, which is particularly important for video applications. In PURSUIT, adding or dropping a multicast subscriber from a group required communication with the RV and TM functions. POINT takes advantage of the fact that the forwarding scheme allows merging unicast paths into multicast trees by using a simple bitwise operation over the FIDs of the unicast paths; thus, multicast senders cache unicast paths to receivers and combine them dynamically into multicast trees as needed. As cached paths may need to be invalidated when the topology changes, POINT developed a network monitoring scheme that allows network changes to be quickly communicated to all interested parties, thus allowing paths to be recalculated only when needed.

To preserve the IP interface towards UEs and other operators, POINT uses a gateway approach. The NAPs, which are the access gateways of customers to the network, or the border gateways to peering networks, handle all the protocols offered at the IP interface, either directly at the IP layer, or, if possible, at the transport or application layer, for example, HTTP. As a result, the POINT network looks like a standard IP network to UEs and peering networks. 

\section{Implementation of IPTV services}

Many ISPs offer IPTV services, consisting of a set of live TV channels delivered via the Real-time Transport Protocol~(RTP) over UDP and IP multicast. The transmitter addresses each TV channel to a separate IP multicast address and receivers subscribe to the multicast group corresponding to the TV channel they want to watch. As IPTV services rely on partially non-standard hardware and software,  we will base our description on the IPTV service offered by PrimeTel in Cyprus, which was also used for the POINT trial. 

\begin{figure*}
\centering
\includegraphics[width=6.5in]{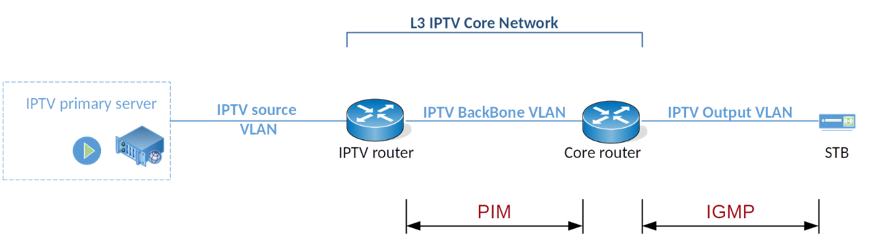}
\caption{PrimeTel's IPTV logical network topology.}\label{iptv}
\end{figure*}

\subsection{IPTV over IP multicast}\label{snoop}

In the PrimeTel IPTV service, whose logical layout is depicted in Figure~\ref{iptv}, the video/audio streams (TV channels) produced by the IPTV server are sent to the IPTV Core Network using a separate IP Multicast address per stream. The IPTV Core Network, an L3 (routed) network, uses Protocol Independent Multicast~(PIM) to route these streams between core routers. The video streams exit the IPTV Core Network and enter the Metro Network, an L2 (switched) Network, which, in turn, feeds the Access Networks (DSLAM, DSL Modem and Set Top Box, or STB). The Metro and Access networks use the Internet Group Management Protocol (IGMP)~\cite{igmpv2}\footnote{We assume IGMP version 2; the extensions for IGMP version 3 are trivial.} to signal the groups that the network should deliver to each user. The STBs receive commands from the user's remote control and issue IGMP ``join'' and ``leave'' messages to indicate which channel the user wants to receive: a channel switch translates to an IGMP ``leave'' for the previous channel and an IGMP ``join'' for the new channel. The video is decoded by the STB and shown on an attached screen.

To avoid forwarding all groups to all customers, the L2 switches at the Metro and Access Networks use \emph{IGMP snooping}~\cite{snoop} to read the L3 headers of the packets and detect IGMP messages. Without IGMP snooping, the switch would flood multicast packets to every port, which would result in broadcasting each group to all end hosts. With IGMP snooping, the switch ``listens'' for the exchanges of IGMP messages between the router and the end hosts and builds a list of all the ports that have requested a particular multicast group. When an IGMP ``join'' message arrives for a stream over port $n$, the stream starts being forwarded over the port; when an IGMP ``leave'' message arrives for that stream over port $n$, the stream stops being forwarded over that port. Due to IGMP snooping, only one stream per channel reaches each L2 switch, therefore the switch is responsible for copying and forwarding the stream to the ports needed. As a result, for each IPTV stream, a multicast tree is formed with an IPTV Core router as the root and the STBs as the leaves. 

\subsection{IPTV over POINT}

In the trial, the POINT network replaced part of the Metro and Access Networks of PrimeTel, an L2 network of IGMP snooping switches. Although extending POINT to replace the entire multicast routing substrate is straightforward, the L2 network is the most interesting part, as it relies on non-standard extensions to IP (IGMP snooping) and the specific topology of the network (tree) to efficiently support multicast. In order for the POINT platform to support the traffic between the unmodified IPTV Server and STBs, the NAPs must translate IP into ICN semantics and vice versa, handling the IP Multicast packets sent from the IPTV Server to the STBs and the IGMP messages exchanged between them. Essentially, the POINT network acts as a ``big'' L2 IGMP Snooping switch, forwarding the appropriate IPTV streams from the server-side NAP (sNAP) connected to the IPTV server, to the L2 devices connected to the client-side NAPs (cNAPs) leading to the STBs.

\begin{figure}
\centering
\includegraphics[width=3.25in]{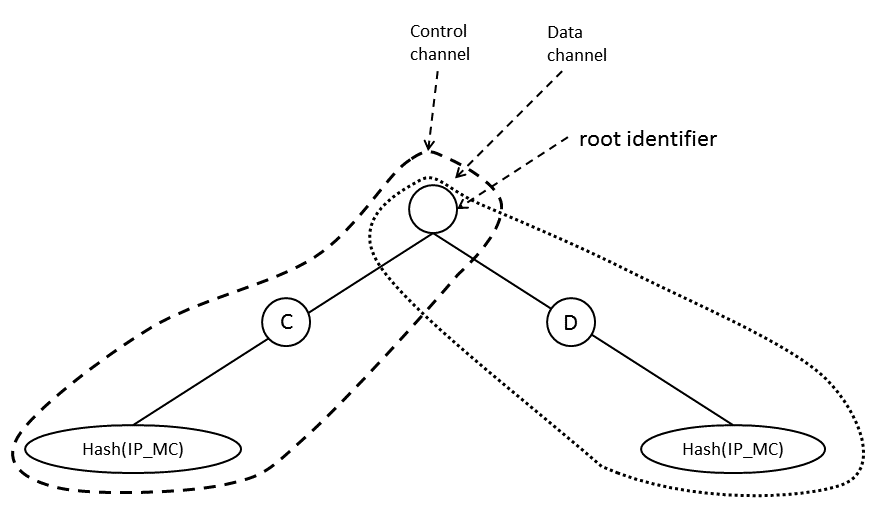}
\caption{Mapping IP multicast groups to names.}\label{ipm}
\end{figure}

Our solution realizes the IGMP operations for joining and leaving an IP multicast group via publications to an appropriate ICN name, representing a control channel, while sending data to the group is realized via publications to another appropriate ICN name, representing a data channel. Figure~\ref{ipm} shows the ICN namespace used for IP multicast. A single root identifier (e.g., $IPMoverICN$) is used for all IP multicast over ICN communication, with two sub-names for the IGMP control ($C$) and the IP multicast data ($D$) channels and a flat namespace of control and data channel names below them. For example, a group with address $IP\_MC$ would map to the $/IPMoverICN/C/Hash(IP\_MC)$ control channel and the $/IPMoverICN/D/Hash(IP\_MC)$ data channel.

A sender to an IP multicast group acts as a subscriber to the corresponding control channel, in order to be notified about join and leave messages to that group. On the other hand, the receivers of the IP multicast group act as subscribers to the corresponding data channel, in order to receive data packets. When a join or leave message is sent by a receiver, it is translated to a publication to the control channel with an implicit subscription or unsubscription to the data channel. With implicit (un)subscriptions, there is no need to involve the RV and TM in each operation: the sNAP simply ORs the FIDs of the unicast paths to each cNAP \emph{currently} participating in the group to form a multicast tree. There is no need to maintain per group state inside the network, as all forwarding is based on FIDs selected at the edges, exactly as in unicast. The only state maintained is at the cNAPs and sNAP.

We can now explain how each IP multicast operation is realized in POINT by the NAPs. We use a static configuration file at each sNAP showing which IPTV channels (or, IP multicast groups) will originate from attached servers. This allows the sNAP to subscribe to the control channels corresponding to the appropriate IP multicast group. There is no need for such configuration at the cNAPs: they will learn which IPTV channels are needed from the IGMP messages sent by the attached IP UEs.

In order for an IP UE to join an IP multicast group as a receiver, it sends an IGMP membership report towards its local cNAP. Upon receiving this, the cNAP checks if the group is already being received. If not, the cNAP sends an implicit subscription message to the corresponding control channel; this is received by the sNAP subscribed to that channel on behalf of the corresponding IP multicast sender. On the sNAP side, we also check if the cNAP is already receiving this group. If so, the message is ignored as a duplicate; if not, then the cNAP is added to the recipients of the group. The underlying ICN system recalculates automatically the FID for the multicast tree whenever such changes occur, allowing IP packets to be delivered via native multicast to the correct cNAPs. 

In order for an IP UE to leave an IP multicast group, it sends an IGMP leave. Upon receiving a leave, the cNAP checks if the group is present. If yes, the cNAP sends a number of IGMP group specific queries for that group. If another IGMP membership report arrives for the group, the leave request is ignored. Otherwise, the cNAP deletes the group and sends an implicit unsubscription message to the corresponding control channel. On the sNAP side, we check if that cNAP is subscribed to that group. If not, the request is ignored; otherwise, the cNAP is removed from the receivers of the group and the FID used for the remaining recipients is recalculated locally.

IPTV servers simply send packets to the IP multicast group via their sNAP. Upon receiving an IP multicast packet, the sNAP checks if there are cNAPs interested in the group. If so, the sNAP publishes the packet under the group name corresponding to its IP address. The receiving cNAPs simply forward the message over their local network using regular IP multicast.

Note that different groups can use different sNAPs as their entry points to the network, by simply configuring each sNAP to subscribe to the corresponding control channel names. When a cNAP sends a join or a leave request by publishing it to the control channel, the proper sNAP receives it and acts accordingly.

\subsection{Handling exceptional conditions}

A common exceptional condition in operational networks is path failure, due to link, switch or router failures. Networks are therefore built with redundant links and nodes in order to avoid interrupting services when such events occur. As an example, Figure~\ref{topology} shows two switches connected by redundant links, a primary and a backup. 

In an existing IPTV network, when there are multiple paths between switches, a spanning tree is created over the topology and some links are disabled for traffic forwarding purposes in order to avoid loops. When IGMP snooping switches look inside IGMP messages, they essentially learn their position in the tree, that is, where IP multicast messages are coming from (the port leading to the IPTV server) and where they should be propagated to (the ports leading to the STBs that have subscribed to this group).  

When a link on the spanning tree fails, the switches attached to it trigger re-convergence, following the recalculation of the spanning tree. Due to the topological change however, the multicast forwarding tree information gathered by the IGMP snooping switches is now partially invalid. As a result, the switches stop forwarding traffic for some time, until re-convergence occurs and new IGMP messages are received, allowing the switches to recreate the multicast tree. Therefore, during link failover, we expect a short service disruption in the order of seconds (depending on the capabilities of the L2 switches: less than a second for fast-converging switches and up to a few seconds for switches without this feature) until the spanning tree is re-created, the switches converge and reports are received from all active STBs.

On the other hand, with POINT and the seamless integration of SDN to the content forwarding process, the link failure is addressed on the data plane, without any interaction with control plane elements. When a failure occurs on the main link between the two switches, the ingress switch detects the failure and automatically switches traffic to the secondary interface which is configured as part of a failover group with the primary interface, without interacting with the SDN controller or any POINT component. No recalculation of multicast trees or FIDs is needed. Therefore, very low packet loss and service disruption is expected, compared to the legacy IP-based IPTV network. 


\section{The closed POINT trial}

\subsection{Overview}

From the outset of the POINT project, the goal was to test the prototype platform produced by the project in a trial, conducted in the operational network of PrimeTel in Cyprus, so as to test the POINT prototype, refine it for operational deployment and exhibit its potential in a real ISP environment. A closed trial was concluded in late 2017 at the headquarters of PrimeTel, using the operational network of the company and its actual IPTV service, with participants who volunteered to test the service.

As part of the trial, users were asked to watch live television channels served over IPTV. The content first travelled over a traditional IP network and then over a POINT-enabled network. During the test, we applied exceptional conditions simulating a link being broken and then repaired in the network. In addition to traditional techniques, such as interviews and questionnaires, to gather user responses, we also used EEG (electroencephalogram) headsets to read user brainwave patterns. This enabled us to measure how levels of frustration increase subconsciously when users are faced with the kinds of exceptional conditions they experienced during the trials, and which can be alleviated through the use of POINT.

\subsection{Trial deployment}

\begin{figure}
\centering
\includegraphics[width=3.25in]{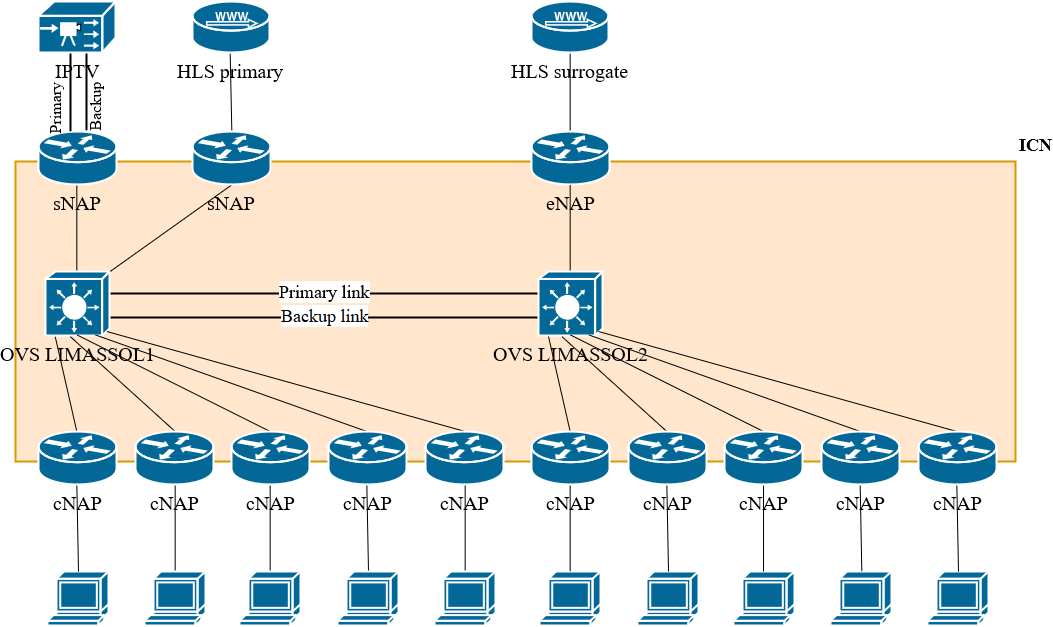}
\caption{Logical trial topology.}\label{topology}
\end{figure}

Figure~\ref{topology} shows the logical topology implemented for the closed trial. The shaded area in the figure represents the POINT network, which is connected via a set of NAPs to regular IP clients and servers. On the top of the figure there is a server offering PrimeTel's IPTV service over UDP/IP mutlicast connected to a server side NAP~(sNAP).\footnote{The trial also involved HLS-based video services, hence the other servers (HLS primary and HLS surrogate) and NAPs (second sNAP and eNAP) in the figure.} On the bottom we have a number of clients, each connected to a client side NAP~(cNAP) via PrimeTel's production ADSL network. The clients are the regular PrimeTel Set Top Boxes~(STBs) used for IPTV services. The sNAPs and cNAPs are connected to two SDN switches, which are interconnected via two trunk links (primary and backup). We used Open vSwitch for the SDN, controlled by an OpenDaylight controller (not shown). The POINT software prototype used in the NAPs ran on regular Debian Linux 8.

This topology is a simplified version of actual ISP topologies that span multiple cities: in each city there are one or more distribution networks, with a number of customers downstream, and possibly some servers upstream. Normally, customers are served by local servers, if available, but when such servers fail, they are served via the interconnection trunks from servers on other networks. It should be noted that the video servers are the actual production servers used by PrimeTel and the access network is PrimeTel's ADSL network. The core network used for the closed trial is PrimeTel's R\&D network. For comparison purposes, the POINT network is running side-by-side with a regular IP network with the exact same topology, using VLANs to allow both types of network to operate all the way from the production servers to the client devices.




\subsection{Trial execution}

Between November 20th and December 1st 2017, we conducted the closed trial in PrimeTel's offices in Cyprus. More than 30 volunteers participated in the study, which involved viewing videos over different networks, under regular and exceptional conditions, and assessing the results with questionnaires. The participants were first introduced to the trial. Then, participants viewed IPTV-based content first over IP and then over POINT, with exceptional conditions occurring during each test. Afterwards, they completed questionnaires to assess their experience. A final exit interview was conducted, before concluding the trial. During the trial, the network was monitored, gathering a wealth of information.

Since the performance of PrimeTel's network and services is already of production quality, under normal operating conditions POINT simply had to match this behavior; indeed, the closed trial did verify that under regular operation, the services provided over POINT were indistinguishable from those provided over IP. The objective of the closed trial was rather to demonstrate that under exceptional network conditions POINT can result in a \emph{better} experience for the viewer. We subjected viewers into exceptional conditions with IPTV, both over the IP and the POINT network, and assessed both the objective performance of the network and the subjective evaluation of the service by the users. All sessions were recorded on video and various interesting events (e.g., video artifacts, noticeable viewer behavior, etc.) were logged. For the IPTV service, the exceptional condition was link failure between the switches serving the customers. In both the POINT and IP case, there are two links between the switches (see Figure~\ref{topology}), but while in the IP case the spanning tree protocol uses only one link, blocking the second to be used as a backup, in POINT both links are active and available all the time. 

\subsection{QoS evaluation}

\begin{figure}
\centering
\includegraphics[width=3.25in]{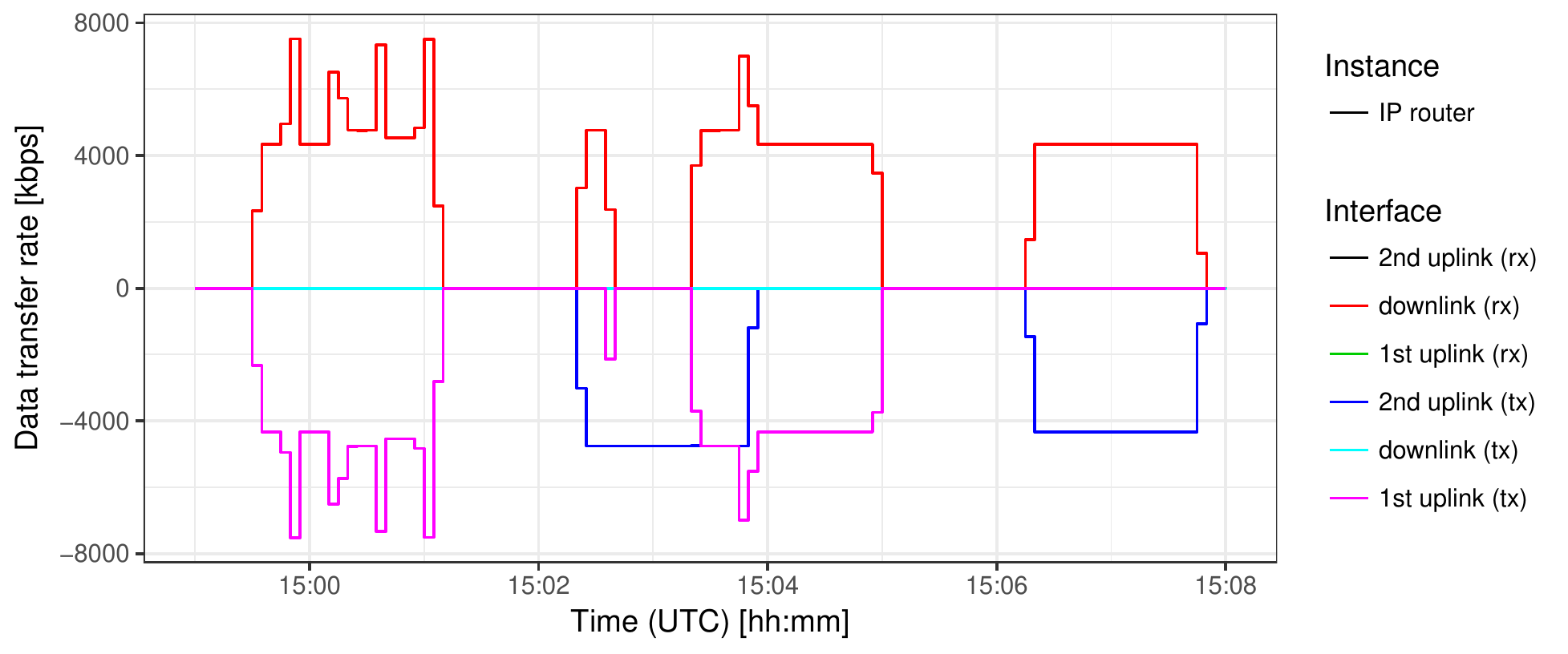}
\caption{Effect of link failover with IP.}\label{ip}
\end{figure}

In the IP case, we brought down the primary interface, which led to recalculation of the spanning tree and re-establishment of the IGMP snooping state, causing major viewing disruption. When the primary interface was brought back up, the same steps were repeated, leading to another service disruption. This can be seen in Figure~\ref{ip}, which shows the data transfer rates of the two uplinks from the IPTV server (bottom part) and of the downlink to the STB (top part). The user's viewing session begins with the first uplink being active (purple line). Approximately at 15:01, the link is disconnected, which triggers the recalculation of the spanning tree and the re-establishment of multicast state, resulting in the user's video freezing (red line drops to zero). Following convergence, the second uplink becomes active (blue line) and the video resumes at around 15:02:10. About 10 sec later, the first link is re-connected, which triggers another re-convergence event and video freezing (red line drops to zero again). The first uplink becomes active at 15:03:20 and the user's viewing resumes normally. Finally, another uplink disconnection at 15:05 resulted in another viewing disruption which was fixed at 15:06:10 through the aforementioned procedure.

\begin{figure}
\centering
\includegraphics[width=3.25in]{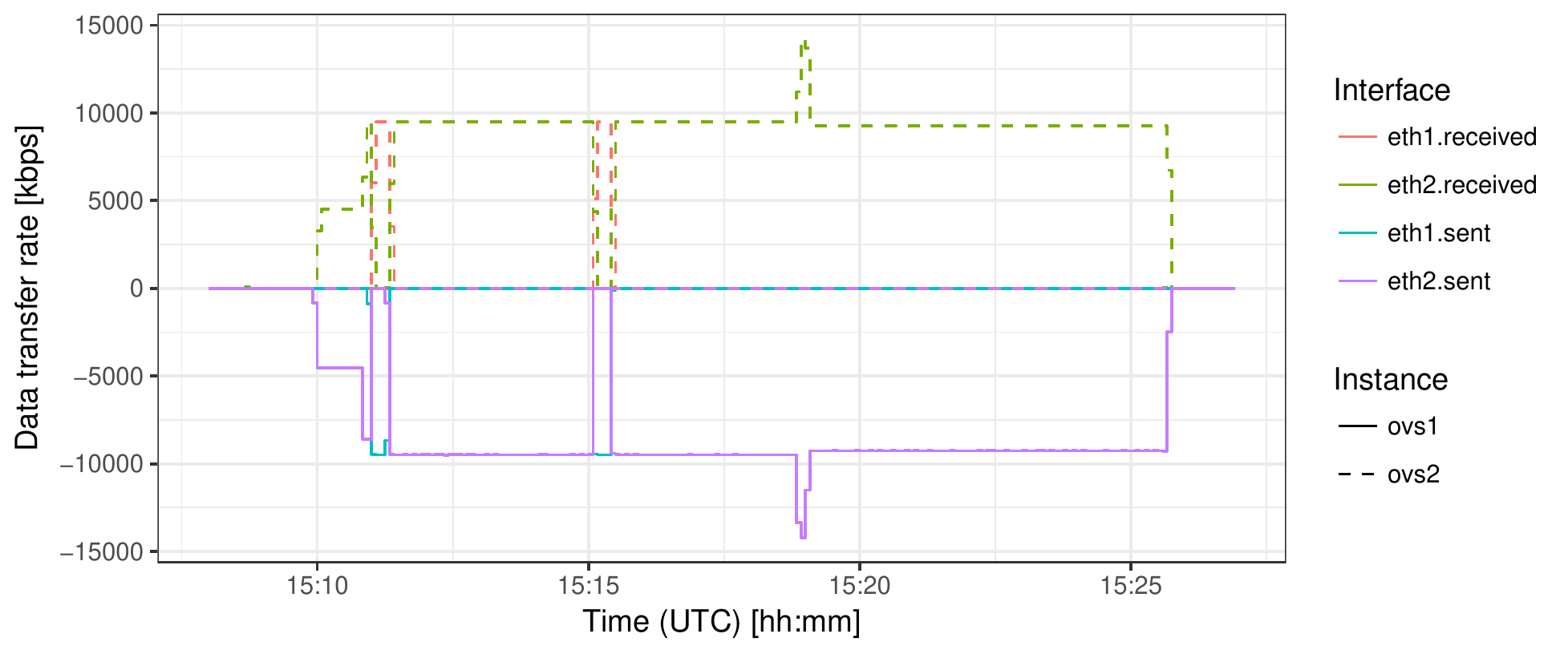}
\caption{Effect of link failover with POINT.}\label{sdn}
\end{figure}

In the ICN case, the failure of the primary link led to seamless switchover to the backup link, while bringing back the primary link led to another seamless switchover to that; in both cases, there were no noticeable disruptions in the service. This can be seen in Figure~\ref{sdn}, which shows the data transmitted by the first OVS instance (ovs1, bottom part) in Figure~\ref{topology} and received by the second OVS instance (ovs2, top part) while the same user as before was viewing IPTV over POINT. Failure of the primary link (eth2), which lasted around 20 sec, and fallback to the backup link (eth1) are visible at 15:11 and 15:15. In contrast to Figure~\ref{ip} and IP, recovery was almost instant and the user's viewing session was not disrupted in any way.

\subsection{QoE evaluation}

Using an adapted version of the i-QoE questionnaire~\cite{i-qoe} we assessed the perceived QoE of IPTV users. The questionnaire includes eighteen items designed to measure user QoE in terms of satisfaction, involvement, enjoyment, endurability and perceived visual quality. Two of the items (9 and 18) use a five-point Likert-scale, while the rest use a seven-point Likert scale. To investigate the impact of network type (POINT vs. IP) on specific QoE questionnaire items, a series of one-way ANOVAs were conducted, taking each questionnaire item as a dependent variable and the network type as the independent variable. The network type has a strong impact on:
\begin{itemize}
\item Item 10 (``I was satisfied with the experience in watching this video clip'') (F (1, 76) = 57.357; p $<$ .05))
\item Item 11 (``I was pleased with the experience in watching this video clip'') (F (1, 76) = 69.024; p $<$ .05))
\item Item 12 (``I was contented with the experience in watching this video clip'') (F (1,76) = 55.376; $<$ .05)) 
\item Item 18 (``Did you perceive any visual impairments in the video (e.g., blockiness, blur, ringing'') (F (1, 76) = 90.652; p $<$ .05))
\end{itemize}

\begin{table}
\centering
\caption{QoE questionnaire items affected by network type.}\label{qtab}
\begin{tabular}{lcccc} \hline 
Item & POINT & Std. Dev.	& IP &  Std. Dev. \\ \hline
Item 10&	5.72	&0.82	&3.51	&1.62\\ \hline
Item 11&	5.72	&0.99	&3.23	&1.58\\ \hline
Item 12&	5.48	&1.02	&3.25	&1.56\\ \hline
Item 18&	4.35	&0.81	&2.00	&1.31\\ \hline
\end{tabular}
\end{table}

The result for Item 10 (see Table~\ref{qtab}) suggests that participants rated their IPTV viewing experience as more satisfying when the content was delivered over POINT compared to when it was delivered over IP. The pattern of strongly positive viewing experiences with video content over  POINT is repeated in Items 11 and 12 . It is evident that all participants found the experience of watching digital video more pleasing and felt more contented when the IPTV content was delivered over POINT. Finally, the responses on Item 18 show that participants thought that any visual impairments when the video was distributed over POINT were imperceptible, while they found the visual impairments when the video was distributed over IP annoying. As a result, their viewing experience was more satisfying (see items 10 -- 12) when video was distributed over POINT as opposed to IP. The interviews follow the same pattern as the questionnaires. Overall, participants thought that their viewing experiences with POINT were much better compared to IP. 

\begin{figure}
\centering
\includegraphics[width=3.25in]{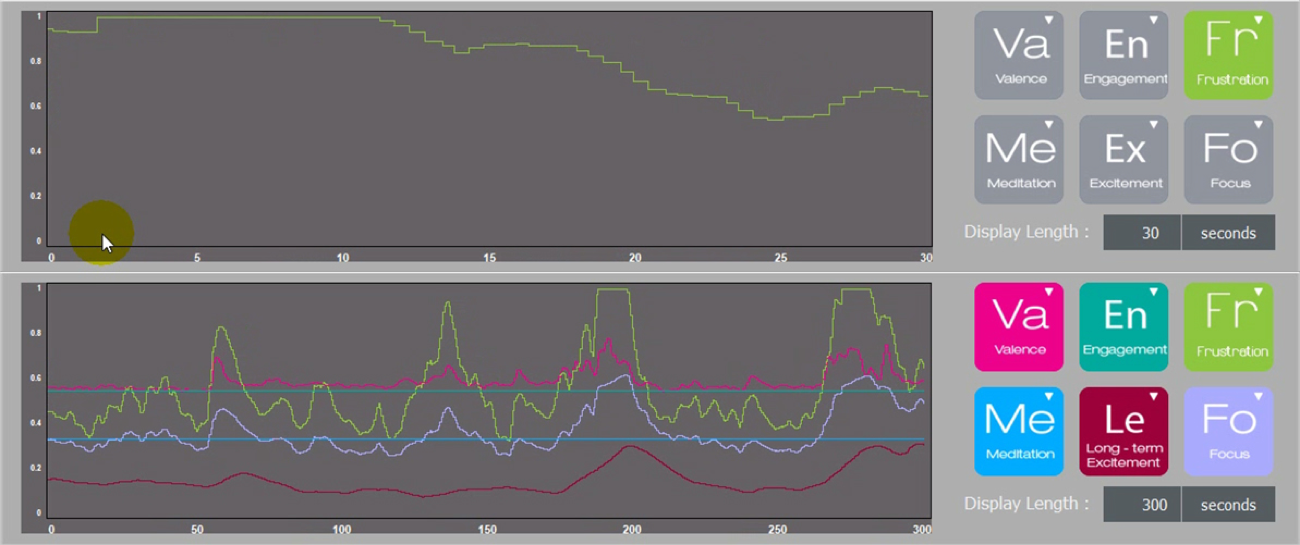}
\caption{Sample output of the EEG analyser.}\label{eeg}
\end{figure}

We also conducted a study with the EEG analyser to record the brain activity of users when viewing IPTV content with both network types. Figure~\ref{eeg} shows the output of the EEG during a test using the IP network, indicating the levels of frustration. We captured a wealth of data (15 million EEG entries from 8 participants in the study) that will take time to analyse properly. An initial review of the data for a single participant shows that there may be some strong emotional responses when the IPTV content was delivered over IP. The data captured from the interviews confirm this finding. All participants said in the interview that their viewing experience with the IPTV content running on IP was a very frustrating experience. Their experience with the IPTV content running on POINT did not produce any negative reactions. 

\section{Conclusion and future work}

Although implementing IP services over ICN is not a novel proposition, POINT is the first project that aims to support any kind of unmodified IP-based service over an actual operator network. The real goal of this approach is to offer improved performance for specific type of traffic, without requiring the entire Internet to transition to ICN. In this paper we explained how POINT can improve the resilience of IPTV services implemented over IP multicast, handling link failures transparently to the users. The closed trial in PrimeTel's network, a first for an ICN project, showed that IPTV-based video services over POINT perform better than over IP under exceptional conditions common in operator networks, while maintaining equal quality during normal operation. 

The final step in the POINT project is to conduct an open trial, which is ongoing. The open trial will take place in actual user homes, using the same equipment and services that PrimeTel uses, and will run for two weeks, allowing us to gather large amounts of operational data on network and service performance. The questionnaires gathered from the participating users will provide additional insights on the QoE offered by POINT over longer periods of time. 

\section*{Acknowledgments}
This research was supported by the EU funded H2020 ICT project POINT under contract 643990, and the RC-AUEB funded ``Original Scientific Publications'' project under contract ER-2766-01.


\end{document}